\documentclass[aip,jcp,reprint]{revtex4-1}
\usepackage{amsmath} 
\usepackage{esint}
\usepackage{graphicx}
\usepackage{color}

\begin{document}

\title{Microscopic analysis of sound attenuation in low-temperature
amorphous solids reveals quantitative importance of non-affine effects} 
\author{Grzegorz Szamel}
\email[Email: ]{grzegorz.szamel@colostate.edu}
\affiliation{Department of Chemistry, Colorado State University,
Fort Collins, Colorado 80523, USA}
\author{Elijah Flenner}
\affiliation{Department of Chemistry, Colorado State University,
Fort Collins, Colorado 80523, USA}

\date{\today}

\begin{abstract}
Sound attenuation in low temperature amorphous solids originates from their  
disordered structure. However, its detailed mechanism is still being debated. 
Here we analyze sound attenuation starting directly from the microscopic equations 
of motion. We derive an exact expression for the zero-temperature sound
damping coefficient. We verify that the sound damping coefficients calculated
from our expression agree very well with results from independent 
simulations of sound
attenuation. The small wavevector analysis of our expression shows that 
sound attenuation is primarily determined by the non-affine displacements' 
contribution to the sound wave propagation coefficient coming from the 
frequency shell of the sound wave. Our expression involves only quantities 
that pertain to solids' static configurations. It can be used to evaluate
the low temperature sound damping coefficients without directly simulating
sound attenuation. 
\end{abstract}

\maketitle

\section{Introduction}

The physics of sound attenuation in amorphous solids is drastically different
than in crystalline solids. 
At low temperatures, when thermal effects can be neglected, sound is attenuated due to 
the inherent disorder of amorphous solids, whereas the attenuation is absent in
crystalline solids. To understand the physical mechanism behind sound attenuation one can
examine its wavevector $k$ dependence.
Sound attenuation in amorphous solids
has a complicated dependence on the wavevector \cite{MasciovecchioPRL2006}, but 
small wavevector $k^4$ scaling of sound damping coefficients has long been 
conjectured on an experimental basis \cite{Zeller,Zaitlin}. An initial interpretation 
was that this small wavevector behavior 
originates from Rayleigh scattering of sound waves from the solid's 
inhomogeneities. Recent computer simulations 
\cite{Mizuno2018,Wang2019,Moriel2019} verified that 
in classical three-dimensional zero-temperature amorphous solids at the smallest
wavevectors sound damping coefficients scale as $k^4$, 
although a logarithmic correction to this 
scaling was also claimed \cite{GelinNatMat2016}. 

The specific physical mechanism of sound attenuation in low temperature amorphous 
solids is still debated. Zeller and Pohl \cite{Zeller} obtained the Rayleigh
scattering law using an ``isotopic scattering''  \cite{Zaitlin} model in which every 
atom of the glass is an independent source of scattering. Several recent experimental 
and simulational results were analyzed within the framework of the fluctuating 
elasticity theory of Schirmacher \cite{SchirmacherEPL,SchirmacherJNCS,Pogna2019}. 
This theory posits that an amorphous solid
can be modeled as a continuous medium with spatially varying elastic constants. 
The inhomogeneity of the elastic constants causes sound scattering and
attenuation. In the limit of the wavelength being much larger than the 
characteristic spatial scale of the inhomogeneity this mechanism is equivalent 
to Rayleigh scattering and the theory predicts that sound damping coefficients scale 
with the wavevector as $k^4$. If the elastic constant variations have
slowly decaying, power-law-like correlations, the theory predicts a logarithmic
correction to Rayleigh scattering \cite{GelinNatMat2016,CuiSM}. 
Other physical approaches, \textit{e.g.} local oscillator 
\cite{SchirmacherJNCS,Buchenau1992,Gurevich1993,Parshin2007,Schober2011} 
and random matrix \cite{GrigeraParisi,ConyuhBeltukov,Ganter2010}
models, can also be used to derive the Rayleigh scattering law. For this reason, 
Rayleigh scaling cannot serve to distinguish between different models 
\cite{SchirmacherJNCS}, and other model predictions must be used to determine 
the mechanism behind sound attenuation. 

Three recent studies came to very different conclusions regarding 
the applicability of the fluctuating elasticity theory for sound attenuation.
First, Caroli and Lema\^{i}tre \cite{CaroliPRL2019} analyzed a version
of the theory derived from microscopic equations of motion. They obtained 
{\em all} the parameters needed to calculate sound attenuation from the theory 
from the same simulations that were used to test the theoretical
predictions. Caroli and Lema\^{i}tre showed that this version of the theory 
underestimates sound damping coefficients by about two orders of magnitude. 

Second, Kapteijns \textit{et al.} \cite{KapteijnsJCP}
analyzed the dependence of sound attenuation in a two-dimensional glass 
on a parameter $\delta$, which
``resembles'' changing the stability of the amorphous solid. 
To calculate the disorder parameter \cite{SchirmacherEPL} of the fluctuating elasticity 
theory they replaced fluctuations of local elastic constants 
(which are used in the theoretical description) by the sample-to-sample fluctuations of
bulk elastic constants. In this way they were able to sidestep the issue of the 
definition of local elastic constants \cite{MizunoPRE2013} and of the correlation 
volume. While Kapteijns \textit{et al.} showed that the disorder parameter and 
the sound damping coefficient have the same dependence on $\delta$, they left 
the calculation of the pre-factor for the scaling for further research. 

Finally, Mahajan and Pica Ciamarra \cite{Mahajan} argued that sound attenuation 
is proportional to the square of the disorder parameter $\gamma$ according to
a version of fluctuating elasticity theory that incorporates an elastic
correlation length \cite{SchirmacherJNCS,SchirmacherCMP}. They relied upon
a relation between the boson peak, the speed of sound, and an elastic correlation
length to show that the speed of sound and the boson peak frequency can be used 
to infer the change of the sound damping coefficient. Again, the magnitude of the 
sound damping coefficient was not addressed.  

The results described above show that it is difficult to distinguish between and
to validate different semi-phenomenological models invoked to describe sound
attenuation in zero-temperature amorphous solids. One of the reasons is that 
most of these approaches involve an adjustable parameter (or parameters) and therefore
are able to predicts trends rather than absolute values of sound damping 
coefficients. For example, neither 
Kapteijns \textit{et al.} nor Mahajan and Pica Ciamarra calculated the values of 
sound damping coefficients (in contrast to Caroli and Lema\^{i}tre), 
but rather investigated the variation 
of the sound attenuation between different glasses. Limited range of glasses that
can be created \textit{in silico} makes it difficult to distinguish between 
trends predicted by different models or different versions of a model.

Our goal is to understand the microscopic origin of the sound 
attenuation. We derive an exact expression for the
sound damping coefficient in terms of quantities that can be calculated from 
static configurations of amorphous solids, without the need to directly simulate 
sound attenuation. Our expression is analogous to well-known Green-Kubo formulae
for transport coefficients \cite{HansenMcDonald}. The latter expressions allow 
one to calculate transport coefficients without explicitly simulating transport 
processes. While both our expression and Green-Kubo formulae 
need to be evaluated numerically, they can also serve as starting points for 
approximate analyses and treatments that can shed light at the validity of
semi-phenomenological models. We hope that the results of one such analysis,
which we present at the end of the paper, can inspire new models or be incorporated 
into the existing ones.  

In Sec. \ref{analysis} we 
start from the microscopic equations of motion for harmonic vibrations.
We derive an exact equation of motion for an auto-correlation function that 
has been used to determine sound attenuation. We identify the self-energy
and show that its real
part reproduces the non-Born contribution to the zero-temperature
wave propagation coefficients. The imaginary part of the 
self-energy is the origin of sound 
attenuation. We show that sound damping coefficients calculated this way 
agree very well with those obtained from direct simulations of sound attenuation
in zero-temperature glasses with different stability. 
In Sec. \ref{smallwavevector} we present the small wavevector expansion
of our expression for the sound damping coefficient. It 
shows that the limiting $k^4$ sound attenuation originates from the same 
physics as the non-Born contribution to the elastic constants and wave propagation
coefficients, \textit{i.e.} from the forces inducing non-affine displacements, 
which appear due 
to the amorphous solids' disordered structure. More precisely, attenuation of the sound 
wave is primarily determined by the contribution to the non-Born part of the wave 
propagation coefficient from a shell of frequencies around the frequency of the
sound wave. 
We thus show the common origin of the renormalization 
of the elastic constants and of sound attenuation. 
In Sec. \ref{planewave} we discuss the results of an approximate evaluation
of our expression for the sound damping coefficient which assumes that the exact
eigenvectors of the Hessian matrix can be replaced by plane waves. These results
allow us to critically evaluate the relation between our exact expression and the
fluctuating elasticity theory. We end the paper with a discussion of our results 
and related descriptions of the sound attenuation. 

\section{Microscopic analysis of sound attenuation}\label{analysis}

We start from microscopic 
equations of motion for small displacements of $N$ 
spherically symmetric particles of unit mass comprising our model amorphous solid, 
\begin{eqnarray}\label{HEOM}
\partial_t^2 \mathbf{u}_i = -\sum_j \mathcal{H}_{ij} \cdot \mathbf{u}_j.
\end{eqnarray}
Here $\mathbf{u}_i$ is the displacement of the $i$th particle from its 
inherent structure (potential energy minimum) position $\mathbf{R}_i$ and 
$\mathcal{H}$ is the Hessian calculated at the inherent structure,
\begin{eqnarray}\label{H}
\mathcal{H}_{ij} = \sum_{l\neq i} 
\frac{\partial^2 V(R_{il})}{\partial \mathbf{R}_i\partial \mathbf{R}_j}
\end{eqnarray}
where $V(r)$ is the pair potential and $\mathcal{H}_{ij}$ is a 3x3 tensor.

To derive an expression for the sound damping coefficient we use a slightly modified
procedure proposed by Gelin \textit{et al.} \cite{GelinNatMat2016}.
We assume that at $t=0$ the particles are displaced from their equilibrium positions
according to 
$\mathbf{u}_i(t=0) = \hat{\mathbf{e}} \exp(-i\mathbf{k}\cdot\mathbf{R}_i)$, 
$\dot{\mathbf{u}}_i(t=0) = 0$,
where $\hat{\mathbf{e}}$ is a unit vector and wavevector $\mathbf{k}$ is
one of the wavevectors allowed by periodic boundary conditions.
We then analyze the time dependence of 
the auto-correlation function of the single-particle
displacement averaged over the whole system,
$C(t) = N^{-1} \sum_i \mathbf{u}_i^*(t=0)\cdot\mathbf{u}_i(t)$.
We anticipate that in the limit of small wavevectors $\mathbf{k}$ 
the auto-correlation function will exhibit damped oscillations, 
$C(t)\propto \cos(v k t)\exp(-\Gamma(k)t/2)$, and we will identify $v$ as the speed 
of sound and $\Gamma(k)$ as the damping coefficient. 

Solving Eqs. (\ref{HEOM}) with our initial conditions  
is equivalent to solving the following equations
\begin{eqnarray}\label{HkEOM}
\partial_t^2 \mathbf{a}_i = -\sum_j \mathcal{H}_{ij}(\mathbf{k}) \cdot \mathbf{a}_j,
\end{eqnarray}
where $\mathcal{H}(\mathbf{k})$ is the wavevector-dependent Hessian,
$\mathcal{H}_{il}(\mathbf{k}) = 
\mathcal{H}_{il} \exp[i\mathbf{k}\cdot\left(\mathbf{R}_i-\mathbf{R}_l\right)]$, 
with initial conditions 
$\mathbf{a}_i(t=0) = \hat{\mathbf{e}}$,
$\dot{\mathbf{a}}_i(t=0) = 0$.
In terms of the new variables, 
$C(t) = N^{-1} \sum_i \mathbf{a}_i(t=0)\cdot\mathbf{a}_i(t)$.

To analyze $C(t)$ we use the standard projection operator 
approach \cite{Zwanzig}. First, we define a scalar
product of two displacement vectors, $\mathbf{a}_i$ and 
$\mathbf{b}_j$, $i,j=1,...,N$, 
$\left<a|b\right> = \sum_i \mathbf{a}_i^*\cdot\mathbf{b}_i$.
Next, we define a unit vector $\left. |1\right>$ with components
$\mathbf{1}_i = N^{-1/2} \hat{\mathbf{e}}$,
and projection operator $\mathcal{P}$ on the unit vector, 
$\mathcal{P} = \left.|1\right>\left<1|\right.$ and
orthogonal projection $\mathcal{Q}$, 
$\mathcal{Q} = \mathcal{I} - \left.|1\right>\left<1|\right.$
where $\mathcal{I}$ is the identity matrix.

Using the projection operator approach we obtain the following expression for the 
Fourier transform $C(\omega) = \int_0^\infty C(t) \exp(i(\omega+i\epsilon))dt$ of
the displacement auto-correlation function,
\begin{eqnarray}\label{Co}
C(\omega) = \frac{i \left(\omega+i\epsilon\right)}
{\left(\omega+i\epsilon\right)^2-\left<1|\mathcal{H}(\mathbf{k})|1\right> 
+ \boldsymbol{\Sigma}(\mathbf{k};\omega)},
\end{eqnarray}
where the self-energy $\boldsymbol{\Sigma}(\mathbf{k};\omega)$ reads
\begin{eqnarray}\label{SEo}
\boldsymbol{\Sigma}(\mathbf{k};\omega) = \left<1\left|\mathcal{H}(\mathbf{k})\mathcal{Q}
\frac{1}{
-(\omega+i\epsilon)^2+\mathcal{Q} \mathcal{H}(\mathbf{k})\mathcal{Q}}
\mathcal{Q} \mathcal{H}(\mathbf{k})\right|1 \right>. 
\nonumber \\
\end{eqnarray}
Equations (\ref{Co}-\ref{SEo}) are exact. While it is straightforward to
calculate $\left<1|\mathcal{H}(\mathbf{k})|1\right>$, evaluation of 
the self-energy requires inversion of a large-dimensional matrix for
each allowed wavevector. To make the numerical effort manageable, in the 
denominator in Eq. (\ref{SEo}) we approximate $\mathcal{H}(\mathbf{k})$ by $\mathcal{H}$.
As argued in Appendix A, 
this approximation does not influence the small wavevector dependence of the
sound damping coefficients.

\begin{figure}
\begin{center}
\includegraphics[width=\columnwidth]{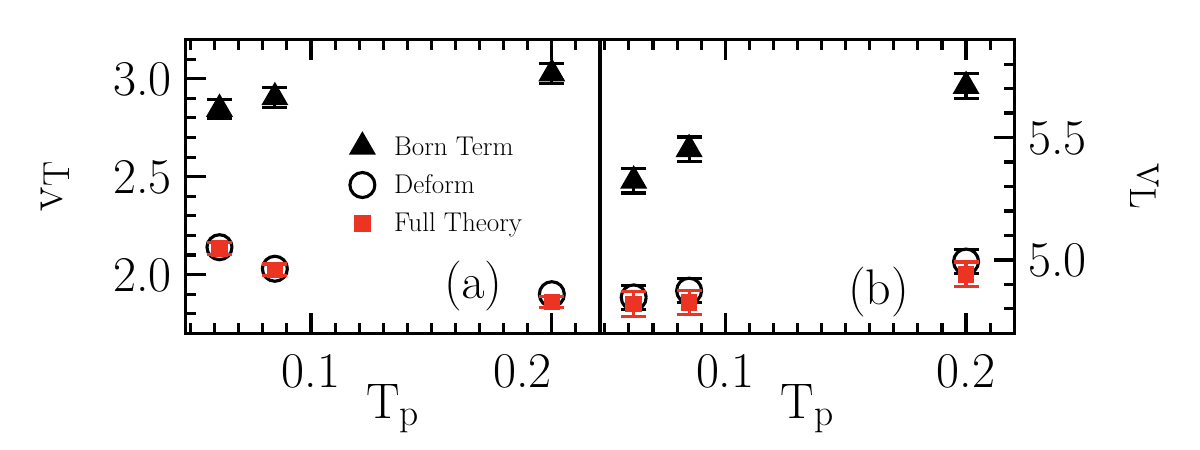}
\includegraphics[width=\columnwidth]{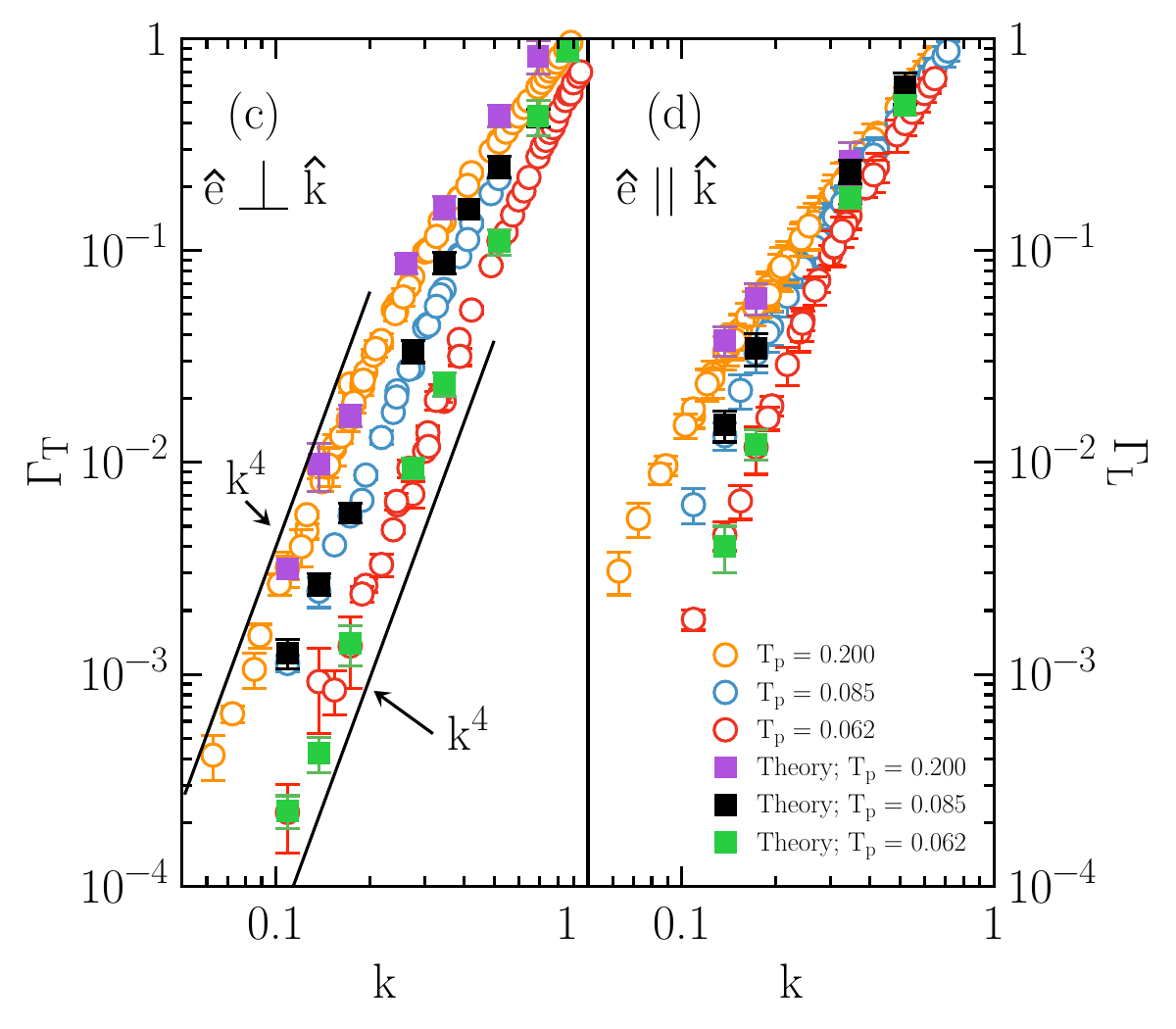}
\caption{\label{fig1} Upper panels: 
the transverse (a) and longitudinal (b) speed of sound 
obtained from the theory, Eq. (\ref{speedr}), 
(red squares) and calculated from the elastic constants (black circles) as a function 
of parent temperature $T_p$. The black triangles are the Born values of 
the speed of sound. Lower panels: the transverse (c) and longitudinal (d) 
damping coefficients obtained from the theory, Eq. (\ref{damping}), (squares) 
and obtained from sound attenuation simulations \cite{Wang2019} (circles) for different 
parent temperatures. Rayleigh scaling $\Gamma \propto k^4$
is recovered at small wavevectors.
The error bars for the theoretical calculation represent the uncertainty due to using 
different bin sizes.}
\end{center}
\end{figure}

In the small wavevector limit, the first non-trivial term in the denominator 
in Eq. (\ref{Co}), $\left<1|\mathcal{H}(\mathbf{k})|1\right>$, can be 
expressed in terms of the Born contributions to the zero-temperature
wave propagation coefficients \cite{Wallace},
\begin{eqnarray}\label{PHkP}
\left<1|\mathcal{H}(\mathbf{k})|1\right> =
\rho^{-1}
\hat{e}_\alpha A^\text{Born}_{\alpha\beta\gamma\delta}
k_\beta \hat{e}_\gamma k_\delta + o(k^2)
\end{eqnarray}
where $\rho=N/V$ is the number density, Greek indices denote Cartesian components, 
the Einstein summation convention for Greek indices is 
hereafter adopted, and $A^\text{Born}_{\alpha\beta\gamma\delta}$
is the Born wave propagation coefficient, which can be expressed 
as the average of the local Born wave propagation coefficients
$A^\text{Born}_{j,\alpha\beta\gamma\delta}$,
\begin{eqnarray}\label{ABornloc}
A^\text{Born}_{j,\alpha\beta\gamma\delta} = \frac{\rho}{2} \sum_{l\neq j} 
\frac{\partial^2 V(R_{jl})}{\partial R_{j,\alpha}\partial R_{j,\gamma}}
R_{jl,\beta}R_{jl,\delta}
\end{eqnarray}
over the whole system, 
\begin{eqnarray}\label{ABorn}
A^\text{Born}_{\alpha\beta\gamma\delta} = 
N^{-1} \sum_j A^\text{Born}_{j,\alpha\beta\gamma\delta} .
\end{eqnarray}
For example, if the coordinate system is chosen such that $\hat{\mathbf{e}}$ is
in the x direction, and we are interested in a transverse wave
and choose $\mathbf{k}$ in the $y$ direction, then the right hand 
side of (\ref{PHkP}) becomes $\rho^{-1} A_{xyxy}^{\text{Born}} k^2$. 

In the absence of the self-energy term, (\ref{Co}) predicts the
Born value of the speed of sound and no sound damping. Both the renormalization
of the sound speed and the sound attenuation originate from the self-energy.

The self-energy can be calculated using the eigenvalues and eigenvectors of the
Hessian. 
In the thermodynamic limit \cite{Mattuck}, when the spectrum of the Hessian
becomes continuous, we can use the Plemelj identity to identify the imaginary 
component of the self-energy, which is responsible for sound attenuation. 
The real $\boldsymbol{\Sigma}'(\mathbf{k};\omega)$ 
and imaginary $\boldsymbol{\Sigma}''(\mathbf{k};\omega)$
parts of the self-energy read,
\begin{eqnarray}\label{SEreal}
\boldsymbol{\Sigma}'(\mathbf{k};\omega) &=& 
\fint d\Omega \, \Upsilon(\mathbf{k},\Omega) \left(\Omega^2-\omega^2\right)^{-1},
\\
\label{SEimag}
\boldsymbol{\Sigma}''(\mathbf{k};\omega) &=& \frac{\pi}{2\omega}
\Upsilon(\mathbf{k},|\omega|),
\end{eqnarray}
where $\fint$ denotes the Cauchy principal value. The function 
$\Upsilon(\mathbf{k},\Omega)$ is defined through the sum over 
eigenvectors $\mathcal{E}^p$ 
of the Hessian matrix with non-zero \cite{commentnonz} 
eigenvalues $\Omega_p^2$ such that 
$\Omega_p\in [\Omega,\Omega+\mathrm{d}\Omega]$, where $\mathrm{d}\Omega$ is the bin size,
\begin{eqnarray}\label{Ups}
\Upsilon(\mathbf{k},\Omega) = (1/\mathrm{d}\Omega)
\sum_{\Omega_p\in [\Omega,\Omega+\mathrm{d}\Omega]}
\left| \left<1\left|\mathcal{H}(\mathbf{k})\mathcal{Q}\right|\mathcal{E}^p\right>
\right|^2.
\end{eqnarray}
The key conceptual issue in writing Eq. (\ref{Ups}) (and closely related equations
(\ref{Th},\ref{De})) is that the thermodynamic limit is implied for the 
expression at the right-hand-side. In this limit the spectrum becomes dense
and phonon bands are not distinguishable. Thus, to calculate 
$\Upsilon(\mathbf{k},\Omega)$ from the analysis of finite-size simulations 
we need to choose bin size $\mathrm{d}\Omega$ such that phonon bands are not
resolved. In the numerical calculations described below we tried a few bin sizes
between 0.1 and 0.2 and found that within this range the results were not very 
sensitive to the bin size.

To evaluate the displacement auto-correlation function we need to find 
complex poles of the denominator at the right-hand-side of Eq. (\ref{Co}). 
In the small wavevector limit this can be done perturbatively, using $k$ as the
small parameter. This leads to the following pair of poles, 
\begin{eqnarray}\label{Copoles}
\omega_\pm = \pm vk - i  \boldsymbol{\Sigma}''(\mathbf{k};vk)/(2vk)
\end{eqnarray}
where the renormalized speed of sound $v$ is given by
\begin{eqnarray}\label{speedr}
v^2 = \lim_{k\to 0} k^{-2} \left[ \left<1|\mathcal{H}(\mathbf{k})|1\right>
-\boldsymbol{\Sigma}'(\mathbf{k};0) \right]  .
\end{eqnarray}
The last term in Eq. (\ref{Copoles}) is our main result. 
It says that the sound damping in zero-temperature
amorphous solids is determined by $\Upsilon(\mathbf{k},\Omega)/\Omega^2$ 
calculated at the wave's frequency, $\Omega=vk$,
\begin{eqnarray}\label{damping}
\Gamma(k) = \frac{\boldsymbol{\Sigma}''(\mathbf{k};vk)}{vk} = 
\frac{\pi}{2} \frac{\Upsilon(\mathbf{k},vk)}{\left(vk\right)^{2}}.
\end{eqnarray}
We emphasize that $\Upsilon(\mathbf{k},\Omega)/\Omega^2$ is the same function that, 
after integration over the whole frequency spectrum,
determines the renormalization of the wave propagation coefficients.
Note that $v$, $\Gamma(k)$, and related quantities defined below
depend on the angle between the polarization of the initial condition
$\hat{\mathbf{e}}$ and the direction of the wavevector $\hat{\mathbf{k}}$.

To verify Eqs. (\ref{speedr}-\ref{damping}) we calculated $v$ and $\Gamma(k)$ for model 
zero-temperature glasses analyzed in Ref.~\cite{Wang2019}. 
These glasses were obtained by instantaneously quenching supercooled liquids 
equilibrated using the swap Monte Carlo algorithm \cite{berthier_prx2017} 
at different parent temperatures $T_p$ 
to their inherent structures 
using the fast inertial relaxation engine 
minimization \cite{method_only_fire}. 
The glasses consist of spherically symmetric, polydisperse particles which
interact via 
a potential $\propto r^{-12}$, with a smooth cutoff, see Appendix B and 
Refs. \cite{Wang2019,berthier_prx2017} for details. The parent temperature controls
the glass's stability and thus its properties \cite{WangNC2019,yielding,Wang2019}.

We calculated eigenvalues and eigenvectors of the Hessian using ARPACK~\cite{arpack}
and Intel Math Kernel Library~\cite{mkl}. Then, using 
Eqs. (\ref{speedr}-\ref{damping}) we evaluated $v$ and $\Gamma(k)$
for the longitudinal, $\hat{\mathbf{e}}\parallel \hat{\mathbf{k}}$, 
and the transverse, $\hat{\mathbf{e}}\perp \hat{\mathbf{k}}$, sound. 
To calculate $\Upsilon(\mathbf{k},\Omega)$ one chooses a $\mathbf{k}_n$ compatible
with periodic boundary conditions. Then one calculates 
$\left|\left<1|\mathcal{H}(\mathbf{k}_n)\mathcal{Q}|
\mathcal{E}^p(\Omega_p)\right>\right|$ 
and bins the results according to the square root of the eigenvalue of 
$\left|\mathcal{E}^p\right>$ to determine $\Upsilon(\mathbf{k}_n,\Omega)$. 
Note that $\Omega_p^2$ is the eigenvalue corresponding to $\left|\mathcal{E}^p\right>$. 
The damping is given by (\ref{damping}) where $\Upsilon(\mathbf{k}_n,\Omega)$
is evaluated at $\Omega = |\mathbf{k}_n|v$. 

Fig.~\ref{fig1} shows results for $v$ and $\Gamma$ 
for three parent temperatures. $T_p=0.2$; 
glasses obtained by 
quenching liquid samples equilibrated at $0.2$ are much less stable than
typical laboratory glasses.   
$T_p=0.085$, which is
between the  mode-coupling temperature $T_c\approx .108$ and the estimated laboratory
glass transition temperature $T_g\approx 0.072$; glasses obtained by 
quenching samples equilibrated at $0.085$ are about as stable as
typical laboratory glasses. $T_p=0.062$, which is well below estimated $T_g$;
glasses obtained by 
quenching liquid samples equilibrated at $0.062$ are as stable as laboratory
ultrastable glasses obtained by the vapor deposition 
method \cite{EdigerScience,EdigerJCP}. 
We previously showed \cite{Wang2019} that sound damping coefficients decrease by
more than an order of magnitude over this range of stability. 

For all three parent temperatures there is excellent agreement between 
results of Eqs. (\ref{speedr}-\ref{damping}) and 
transverse and longitudinal sound speeds, $v_T$ and $v_L$, 
and transverse and longitudinal sound damping coefficients, $\Gamma_T$ and $\Gamma_L$
obtained previously \cite{Wang2019} from direct simulations of sound attenuation.
At small wavevectors we recover Rayleigh scaling, $\Gamma \propto k^4$,
but the theory also accurately predicts sound damping for 
wavevectors outside the Rayleigh scaling regime.
The predicted damping coefficients depart from the 
simulation results for larger wavevectors, but at larger 
wavevectors the assumptions used to find the poles, Eq. (\ref{Copoles}), 
become invalid. 

\section{The origin of sound attenuation: non-affine effects}\label{smallwavevector}

To get some physical insight into the origin of sound attenuation in zero-temperature
amorphous solids we examine the small wavevector expansion of 
$\left<1\left|\mathcal{H}(\mathbf{k})\mathcal{Q}\right|\mathcal{E}^p\right>$, 
\begin{eqnarray}\label{PHkQ}
&& 
\left<1\left|\mathcal{H}(\mathbf{k})\mathcal{Q}\right|\mathcal{E}^p\right> = 
-iN^{-1/2} \sum_j \boldsymbol{\Xi}_{j, \gamma\delta}\hat{e}_\gamma k_\delta\cdot
\boldsymbol{\mathcal{E}}_{j}^p
\\ \nonumber && 
+
\rho^{-1}N^{-1/2}\sum_j \left[A^\text{Born}_{j, \alpha\beta\gamma\delta}
-\hat{e}_\alpha \hat{e}_\mu A^\text{Born}_{\mu\beta\gamma\delta}\right]\hat{e}_\gamma
k_\beta k_\delta \mathcal{E}_{j,\alpha}^p 
+ o(k^2). 
\end{eqnarray}
In Eq. (\ref{PHkQ}) $\boldsymbol{\mathcal{E}}_{j}^p$ denotes the component of the $p$th 
eigenvector of the Hessian corresponding to particle $j$ and $\mathcal{E}_{j,\alpha}^p$ 
denotes its Cartesian component $\alpha$. 
Furthermore, $\boldsymbol{\Xi}_{j,\beta\gamma}$ 
denotes the vector field describing forces due affine deformations,
\begin{eqnarray}\label{Xi}
\boldsymbol{\Xi}_{j,\gamma\delta} = 
-\sum_{l\neq j} \frac{\partial^2 V(R_{jl})}{\partial R_{j,\gamma}\partial \mathbf{R}_{j}}
R_{jl,\delta}.
\end{eqnarray}  
Specifically, $\boldsymbol{\Xi}_{j,\gamma\delta}$ is proportional 
to the force on particle $j$ 
resulting from a deformation along the $\gamma$ direction that linearly depends
on the $\delta$ coordinates. 
Finally, the 2nd term at the right-hand-side of 
Eq. (\ref{PHkQ}) accounts for the spatial variation 
of the local Born wave propagation coefficients.

As discussed in the literature \cite{MaloneyPRL,LemaitreJSP}, 
forces encoded in vector field $\boldsymbol{\Xi}_{j,\gamma\delta}$ 
do not seem to posses any longer-range correlations. In contrast,
non-affine displacements given by $\mathcal{H}^{-1}\cdot\boldsymbol{\Xi}$
exhibit characteristic vortex-like structures and correlations extending
over many particle diameters \cite{MaloneyPRL,LemaitreJSP,LeonfortePRB,DamartPRB}.
The characteristic length of these correlations determines the minimal length scale
on which a macroscopic elastic approach can be used to describe the response of
amorphous solids \cite{LeonfortePRB}. 
It follows from the combination of Eqs. 
(\ref{SEreal}), (\ref{speedr}) and (\ref{PHkQ}) that the renormalization of
the wave propagation coefficients originates from the first 
term in Eq. (\ref{PHkQ}), 
\begin{eqnarray}\label{speedr2}
\lim_{k\to 0} k^{-2} \boldsymbol{\Sigma}'(\mathbf{k};0) = 
N^{-1} \int d\Omega \, \Theta(\Omega) \Omega^{-2}
\end{eqnarray}
where $\Theta(\Omega)$ is defined analogously to $\Upsilon(\mathbf{k},\Omega)$,
\begin{eqnarray}\label{Th}
\Theta(\Omega) = (1/\mathrm{d}\Omega)\sum_{\Omega_p\in [\Omega,\Omega+\mathrm{d}\Omega]} 
\left|\boldsymbol{\Xi}_{\gamma\delta}^p\hat{e}_\gamma \hat{k}_\delta\right|^2
\end{eqnarray}
with 
\begin{eqnarray}\label{Xip}
\boldsymbol{\Xi}_{\gamma\delta}^p = N^{-1/2} \sum_j
\boldsymbol{\Xi}_{j,\gamma\delta}\cdot\boldsymbol{\mathcal{E}}_{j}^p.
\end{eqnarray} 
Equations (\ref{speedr2}-\ref{Th})
reproduce the exact expression for the non-Born contribution to the 
wave propagation coefficients derived from the analysis of the non-affine
displacements \cite{MaloneyPRL,LemaitreJSP}.
We note that function $\Theta(\Omega)$ is closely related to 
function $\Gamma_{\alpha\beta\kappa\chi}(\omega)$ introduced and evaluated
by Lema\^{i}tre and Maloney, see Eq. (32) of Ref. \cite{LemaitreJSP}.

\begin{figure}
\includegraphics[width=\columnwidth]{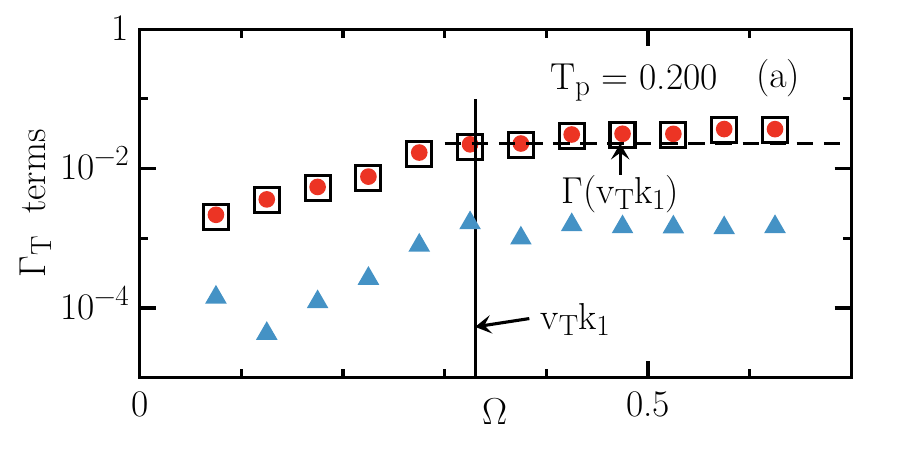}
\includegraphics[width=\columnwidth]{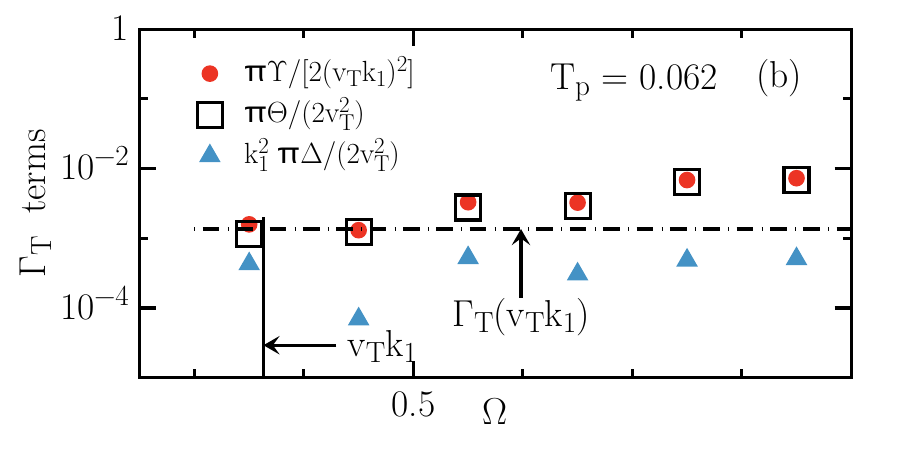}
\caption{\label{Terms}
The terms that contribute to the transverse 
sound attenuation for $k_1 = 2\pi/L = 0.13722$,  where 
$L$ is the length of the simulation box, given by Eq.~\ref{damping2}
for $T_p=0.200$ (a) and $T_p=0.062$ (b). The vertical 
lines marks the frequency $\Omega = v_T k_1$, where $v_T$ is the transverse speed of
sound, and the horizontal line is the sound attenuation 
calculated in simulations from Ref.~\cite{Wang2019}. 
The value of $\pi \Upsilon(\mathbf{k}_1,\Omega=v_T k_1)/[2(v_T k_1)^2]$ 
gives the damping for $k_1$.
}
\end{figure}

While only the first term in Eq. (\ref{PHkQ}) determines the renormalization 
of the wave propagation coefficients,
both terms contribute to sound attenuation,
\begin{eqnarray}\label{damping2}
\Gamma(k) = \frac{\pi}{2v^2}\left[\Theta(vk) + k^2 \Delta(vk)\right]
\end{eqnarray}
where $\Delta(\Omega)$  is defined analogously to $\Upsilon(\mathbf{k},\Omega)$,
\begin{eqnarray}\label{De}
\Delta(\Omega) &=& 
(1/\mathrm{d}\Omega)\sum_{\Omega_p\in [\Omega,\Omega+\mathrm{d}\Omega]} 
\left|\delta A_{\beta\gamma\delta}^p\hat{k}_\beta \hat{e}_\gamma \hat{k}_\delta
\right|^2
\end{eqnarray}
with 
\begin{eqnarray}\label{deltaAp}
\delta A_{\beta\gamma\delta}^p = \rho^{-1} N^{-1/2}
\sum_j \left[A^\text{Born}_{j,\alpha\beta\gamma\delta}
-\hat{e}_\alpha \hat{e}_\mu A^\text{Born}_{\mu\beta\gamma\delta}\right]
\mathcal{E}_{j,\alpha}^p.
\nonumber \\
\end{eqnarray}
We note that the second term in Eq. (\ref{damping2}) is expressed in term of the 
fluctuations of the local Born wave propagation coefficients, see Eq. (\ref{deltaAp}). 
Thus, the physical content of the second term resembles 
that of the fluctuating elasticity theory. We will discuss this correspondence 
further in the next section. 

It is the first term in Eq. (\ref{damping2}) that makes the 
dominant contribution to the damping coefficient, see Fig.~\ref{Terms}.
This implies that the sound damping is primarily determined by 
function $\Theta(\Omega)$, which is the same function that also determines 
the renormalization of the wave propagation coefficients, Eq. (\ref{speedr2}).
While previous studies suggested \cite{CaroliJCP2020} and analyzed approximately 
\cite{BaggioliZaccone} the importance the non-affine effects for the sound attenuation, 
we have presented the first approach that accounts for these effects exactly.

\section{Sound damping in the plane-wave approximation}\label{planewave}

The most recent version of the fluctuating elasticity theory discussed by Mahajan and 
Pica Ciammarra \cite{Mahajan} posits that ``amorphous materials can be described as 
homogeneous isotropic elastic media punctuated by quasilocalized modes acting as elastic 
heterogeneities.'' This suggests that plane waves should be a reasonable zeroth order 
approximation for the eigenvectors of the Hessian matrix describing an amorphous solid.
To check this supposition we calculated $\Theta$ and $\Delta$ contributions in Eq. 
(\ref{damping2}) approximating the exact eigenvectors by plane waves, 
$\boldsymbol{\mathcal{E}}_{j}^p \propto 
\hat{\mathbf{e}}_\mathbf{q} e^{-i\mathbf{q}\cdot\mathbf{R}_j}$, see Appendix C
for details. For the contributions to transverse wave damping coefficient we obtained
the following expressions
\begin{eqnarray}\label{dampingpw1}
&&\frac{\pi}{2v^2}\Theta(v_T k) \approx \frac{1}{60\pi}\frac{k^4}{\rho^3}
\left[\left(\frac{v_T^2}{v_L^5} + \frac{4}{v_T^3}\right)
\left<A^\text{Born}_{\alpha\beta xy}A^\text{Born}_{\alpha\beta xy}\right>
\right. \nonumber \\ && \left. 
+\left(\frac{v_T^2}{v_L^5} - \frac{1}{v_T^3}\right)
\left<A^\text{Born}_{\alpha\alpha xy}A^\text{Born}_{\gamma\gamma xy}
+A^\text{Born}_{\alpha\beta xy}A^\text{Born}_{\beta\alpha xy}\right>
\right],
\end{eqnarray}
\begin{eqnarray}\label{dampingpw2}
&& \frac{\pi k^2 }{2v^2}\Delta(v_T k) \approx
\frac{1}{12\pi}\frac{k^4}{\rho^3}
\left(\frac{1}{v_\text{L}^3}+ \frac{2}{v_\text{T}^3}\right)
\left[\left<\left(\Delta A^\text{Born}_{xyxy}\right)^2\right> 
\right. \nonumber \\ && \left. 
+ \left<\left(A^\text{Born}_{yyxy}\right)^2\right> 
+ \left<\left(A^\text{Born}_{zyxy}\right)^2\right> \right],
\end{eqnarray}
where we implicitly assumed analyticity of the correlation functions of local 
wave propagation coefficients at the vanishing wavevector. 
For example, we assumed that at $\mathbf{q}\to 0$, 
\begin{eqnarray}\label{Bornxyxy}
\left<\left(\Delta A^\text{Born}_{xyxy}\right)^2\right> = 
\lim_{\mathbf{q}\to 0} N^{-1}
\left|\sum_j e^{i\mathbf{q}\cdot\mathbf{R}_j} 
\left(A^\text{Born}_{j,xyxy}- 
A^\text{Born}_{xyxy}\right)\right|^2
\nonumber \\
\end{eqnarray} 
and other similar equalities, as discussed in Appendix C.

We note that while the exact formula (\ref{Th}) for $\Theta$ contribution involves
non-affine forces $\boldsymbol{\Xi}$, approximate formula (\ref{dampingpw1}) is
expressed in terms of correlations of local wave propagation coefficients.
This follows from the fact that, as shown in Appendix C, for small wavevectors
$\mathbf{q}$
\begin{eqnarray}\label{XitoA}
\sum_j
\boldsymbol{\Xi}_{j,\gamma\delta}\cdot 
\hat{\mathbf{e}}_\mathbf{q} e^{-i\mathbf{q}\cdot\mathbf{R}_j} =
\frac{i}{\rho} \sum_j 
A^\text{Born}_{j,\alpha\beta\gamma\delta}
\hat{e}_\mathbf{q\alpha} q_\beta e^{-i\mathbf{q}\cdot\mathbf{R}_j} + o(q).
\nonumber \\
\end{eqnarray} 
Furthermore, we note that formulae (\ref{dampingpw1}-\ref{dampingpw2}) are 
reminiscent of Zeller and Pohl's ``isotopic scattering'' model in that every atom $j$
is a source of scattering of a plane wave, with the amplitude depending on its local 
wave propagation coefficient $A^\text{Born}_{j,\alpha\beta\gamma\delta}$. 
Importantly, our approximate formulae involve correlations of local 
wave propagation coefficients that vanish at the macroscopic level and thus 
do not appear in the semi-phenomenological
fluctuating elasticity theory, 
\textit{e.g.} $A^\text{Born}_{j,yyxy}$.

\begin{figure}
\includegraphics[width=\columnwidth]{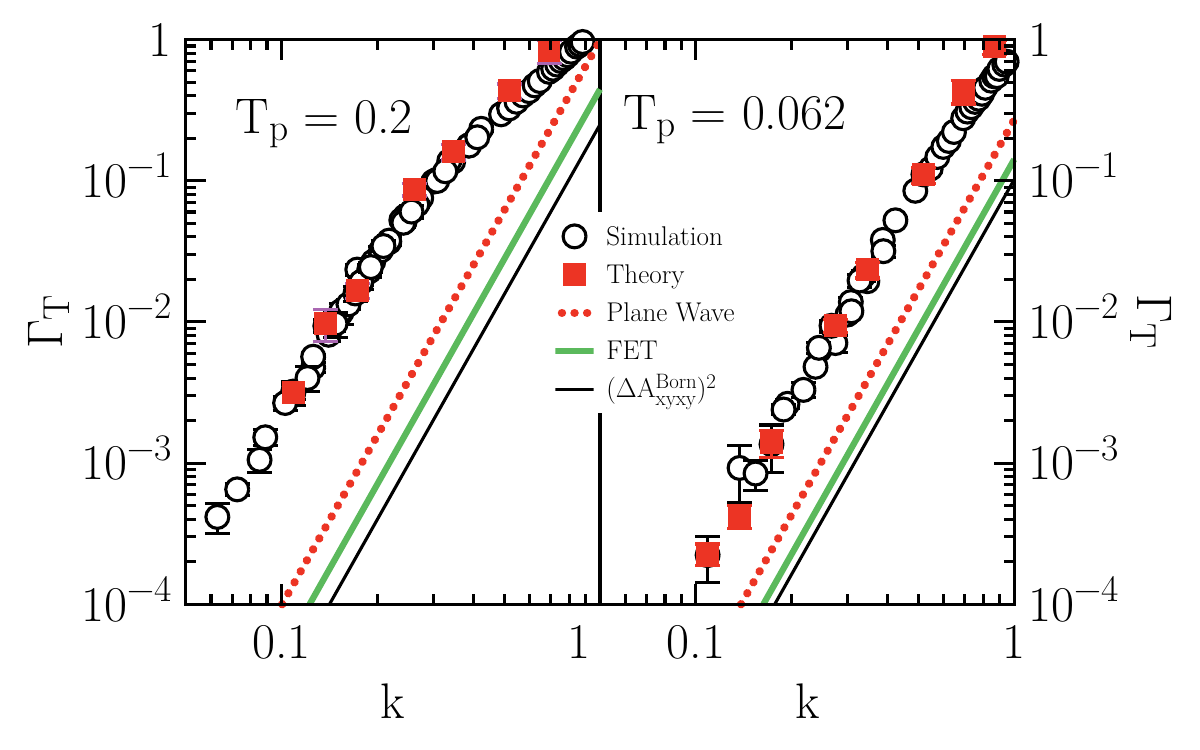}
\caption{\label{Compare} Comparison of the sound damping coefficients obtained
from the theory (squares) and evaluated from sound attenuation simulations (circles) 
with predictions of the plane-wave approximation, 
Eqs. (\ref{dampingpw1}-\ref{dampingpw2}) 
(red symbols) and microscopic version of the 
fluctuating elasticity theory, \textit{i.e.} the 
contribution due to $\left<\left(\Delta A^\text{Born}_{xyxy}\right)^2\right>$ term in
Eq. (\ref{dampingpw2}) (black line). Also shown is the result of 
a semi-phenomenological fluctuating elasticity theory (FET, green line).}
\end{figure}

The plane wave approximation recovers analytically the Rayleigh scattering $k^4$
scaling of the sound damping coefficient. However, it is 
quantitatively quite inaccurate, see Fig. \ref{Compare}. This implies that at least 
for the purpose of calculating sound damping, eigenvectors of the Hessian are not
well approximated by plane waves. We note that the plane-wave approximation
becomes more accurate with decreasing parent temperature or increasing
glass stability.

Finally, we note that the first term in square brackets in Eq. (\ref{dampingpw2}),
which involves correlations of the fluctuations of the local shear modulus, 
$\Delta A^\text{Born}_{j,xyxy}$, represents the result of the microscopic, 
isotopic scattering-like, version of the fluctuating elasticity theory. 
As shown in Fig. \ref{Compare}, this term is about 2.5-4 times smaller
than the complete 
plane-wave result, and thus it severely underestimates sound attenuation. 

In Fig. \ref{Compare} we also show the result of a semi-phenomenological fluctuating 
elasticity theory. To calculate this result we started from the celebrated formula of 
Rayleigh \cite{Rayleigh} that predicts the attenuation of a transverse wave due to 
inclusions of volume $V_d$ and number density $n$,  
$\Gamma^\text{R}(k) = n V_d v_T \gamma k^4/(6\pi)$,
where $\gamma$ is the disorder parameter. 
In Rayleigh's calculation $\gamma$ characterized 
the variation of ``optical density''. To adopt his calculation to the present problem
we expressed $\gamma$ in terms of the variation of the square of the transverse 
speed of sound, $\gamma = (\delta v_T^2)^2 V_d/v_T^4$. Next, we added to Rayleigh's 
expression the the contribution of the longitudinal waves excited due to the presence 
of the inclusions, $\delta \Gamma(k) = n V_d v_T \gamma k^4 (v_T^3/(2 v_L^3))/(6\pi)$. 
The complete formula of the semi-phenomenological fluctuating elasticity theory
thus reads
\begin{eqnarray}\label{FET}
\Gamma^\text{FET}(k) 
= \frac{k^4 v_T}{6\pi} \left(1+ \frac{1}{2}\frac{v_T^3}{v_L^3}\right)
n V_d \gamma .
\end{eqnarray}
We note that if one makes the identification 
$\left<\left(\Delta A^\text{Born}_{xyxy}\right)^2\right>/(\rho^3 v_T^4) = n V_d \gamma$,
the contribution to sound attenuation due to the first term in square brackets in 
Eq. (\ref{dampingpw2}) becomes identical to expression (\ref{FET}). 
To calculate the value of $\Gamma^\text{FET}$ we need the disorder parameter $\gamma$ and
the volume fraction of the inclusions $nV_d$. For $\gamma$ we used previously
obtained results for the fluctuations of local elastic constants \cite{Shakerpoor}. 
We recall that disorder parameters calculated this way increase slightly with increasing
box size used to define local elastic constants, thus we used the largest box size 
considered in Ref.~\cite{Shakerpoor}. Furthermore, we note that 
Mahajan and Pica Ciamarra's 
formulation of fluctuating elasticity theory assumes $n V_d\ll 1$, see the SI of Ref. 
\cite{Mahajan}. To calculate the upper bound for $\Gamma^\text{FET}$ 
we substituted $n V_d=1$.
Figure \ref{Compare} shows that the result of this procedure significantly underestimates
sound attenuation. 

We note that in addition to the microscopic version of the fluctuating elasticity
theory, originally derived by Caroli and Lema\^{i}tre and embodied in the first term 
in square brackets in Eq. (\ref{dampingpw2}), and the semi-phenomemonogical 
approach resulting in expression (\ref{FET}) one could compare our results to
predictions of more sophisticated versions of the fluctuating elasticity 
theory, \textit{e.g.} the version relying upon the self consistent Born approximation
\cite{SchirmacherEPL}. This comparison is left for future work. 

\section{Discussion} 

According to our microscopic analysis, sound attenuation
in zero-temperature amorphous solids is primarily determined by internal forces induced
by initial affine displacements of the particles, \textit{i.e.} by the physics 
of non-affine displacement fields. Quantitatively, the
damping coefficient is proportional to the non-affine contribution to
the wave propagation coefficients from the frequency equal to the frequency of the
sound wave. 
It is not trivial that our exact calculation (as opposed to the plane-wave approximation 
discussed in the previous section) reproduces the Rayleigh scaling
of sound damping coefficients. This fact results from the frequency dependence of
$\Theta$ and $\Delta$, which deserves further theoretical study. 

The mechanism of the attenuation revealed by our microscopic analysis was mentioned by 
Caroli and Lema\^{i}tre in Ref.~\cite{CaroliPRL2019}. It was investigated in 
Ref.~\cite{CaroliJCP2020}, where Caroli and Lema\^{i}tre considered separately the 
effects of the long-wavelength, elastic continuum-like, and small-scale,  
primarily non-affine, motions with the small-scale motions being the  
scatterers for the long-wavelength ones.


An earlier study by
Wang, Szamel and Flenner\cite{Wang2019} found a strong correlation between 
the sound attenuation coefficient and the amplitude of the vibrational 
density of states of quasilocalized modes. The latter modes were defined
using a cutoff in the participation ratio, folowing 
Mizuno \textit{et al.}\cite{MizunoPNAS} and Wang \textit{et al.}\cite{WangNC2019}.
We attempted to quantify the relative contributions of the extended and quasi-localized 
modes by separating the contributions of modes with small and large participation
ratio. We did not find convincing evidence for the dominance of 
small participation ratio modes versus larger participation ratio modes.

We note in this context that local oscillator 
models\cite{Buchenau1992,Gurevich1993,Parshin2007,Schober2011} express the sound 
attenuation coefficient in terms of the contributions from localized 
``defects''\cite{LairdSchober,SchoberLaird} 
referred to as ``soft modes''. The formulas derived in these approaches are 
similar to our Eqs. \eqref{damping} and \eqref{damping2}. The details of 
expressions of Refs.\cite{Buchenau1992,Gurevich1993,Parshin2007,Schober2011} and 
our Eqs. \eqref{damping} and \eqref{damping2} differ; 
in particular we express the self-energy in terms of all the 
exact eigenvectors and eigenvalues of the Hessian matrix. In order to evaluate 
the local oscillator based sound attenuation coefficient formulas one needs 
to characterize the properties of the soft modes. In practical 
applications one may parametrize the soft modes' properties and fit the 
parameters to the experimental results. Such a procedure 
was used by Schober\cite{Schober2011}
and resulted in a good agreement between the theory and experiment. 

In view of both the previously found\cite{Wang2019} correlation between the 
sound attenuation coefficient and the amplitude of the vibrational density of states
of quasilocalized modes and the success of local oscillator approach\cite{Schober2011} 
we believe that future work should investigate whether 
dominant contributions to the sound atteanuation coefficient formulas 
\eqref{damping} and \eqref{damping2} originate from well defined regions 
that can be identified as ``defects''. 

Damart \textit{et al.} \cite{DamartPRB} demonstrated that the non-affine displacement 
field was responsible for high-frequency harmonic dissipation in a simulated
amorphous SiO$_2$. Therefore, it appears that non-affine displacements are
responsible for dissipation over the full frequency range. Further theoretical 
development is needed to connect the low-frequency and high-frequency theories. 

Recently, Baggioli and Zaccone developed an approximate microscopic theory
for the sound attenuation that takes into account non-affine displacements 
\cite{BaggioliZaccone}. This theory shares physical insight with our approach 
but it is quantitatively as inaccurate as the plane-wave version of our exact formula.

As we mentioned in the introduction, Gelin \textit{et al.} \cite{GelinNatMat2016}
found a logarithmic correction to the Rayleigh scattering scaling of
the sound damping coefficients, which within the fluctuating elasticity
theory could originate from the slowly decaying correlations between local
values of the elastic constants \cite{GelinNatMat2016,CuiSM}. Within our approach,
a logarithmic correction could originate from a logarithmic dependence of 
$\Theta(\Omega)$ or $\Delta(\Omega)$ on frequency $\Omega$. Our present numerical data
are consistent with the absence of such a logarithmic dependence but it would 
be interesting to investigate this issue farther.  

Within the plane-wave approximation a logarithmic correction could result from
a logarithmic small wavevector divergence of the correlation functions 
of local Born wave propagation coefficients. We did not observe such a divergence
but we note that our systems were significantly smaller that those discussed 
in Ref. \cite{GelinNatMat2016}. We note that if the correlation functions of
local wave propagation coefficients are singular, additional terms in the
plane-wave approximation will appear. These terms will originate from the 
anisotropic small wavevector character of the correlation functions of
local wave propagation coefficients.

Our approach arrives at the physical picture of sound attenuation
different from that postulated in the fluctuating elasticity theory.
While the latter theory can predict trends \cite{KapteijnsJCP,Mahajan}, it
is quantitatively very inaccurate, 
as noted earlier by Caroli and Lema\^{i}tre \cite{CaroliPRL2019}. 
Our analysis revealed that the fluctuating elasticity theory misses 
the dominant non-affine effects. In addition, it does not include 
the contributions due to fluctuations of local microscopic wave propagation
coefficients that vanish at the macroscopic level. Most importantly, the fluctuating
elasticity theory uses plane-wave-like picture of sound in low-temperature 
amorphous solids. The comparison of the results obtained using the full 
theoretical expression and adopting the plane-wave approximation, 
shown in Fig. \ref{Compare}, suggests that this
leads to large quantitative discrepancies.

Finally, we note that calculating sound attenuation using 
Eq. (\ref{damping}) or (\ref{damping2}) is somewhat numerically demanding 
but more straightforward than analyzing the time dependence of the 
velocity or displacement auto-correlation functions. The latter analysis 
suffers from large finite-size effects \cite{Wang2019,Bouchbinder2018} 
that make the evaluation of the sound damping coefficients at the smallest wavevectors 
allowed by the periodic boundary conditions difficult. Our approach
offers an attractive alternative way to evaluate sound damping coefficients
of low temperature elastic solids that does not suffer from finite size effects. 

\begin{acknowledgments}
We thank A. Ninarello for generously providing  equilibrated configurations 
at very low parent temperatures and E. Bouchbinder for comments on the manuscript.
We gratefully acknowledge the support
of NSF Grant No.~CHE 1800282. 
\end{acknowledgments}

\section*{Author declarations}

\subsection*{Conflict of interest}

The authors have no conflicts to disclose.

\section*{Data availability}

The data that support the findings of this study are available from 
the corresponding author upon reasonable request.

\appendix
\numberwithin{equation}{section}
\section{Approximation 
$\mathcal{H}(\mathbf{k})\approx \mathcal{H}$ in Eq. (\ref{SEo}) of the main text} 
First, we examine the small wavevector expansion of 
$\mathcal{Q} \mathcal{H}(\mathbf{k})\mathcal{Q}$. The $i,j$ element,
which is a 3x3 tensor, reads
\begin{widetext}
\begin{eqnarray}\label{QHkQ}
&& \left[\mathcal{Q} \mathcal{H}(\mathbf{k})\mathcal{Q}\right]_{ij} = 
\mathcal{H}_{ij}e^{-i\mathbf{k}\cdot\left(\mathbf{R}_i-\mathbf{R}_j\right)}
-N^{-1} \sum_{l} \mathcal{H}_{il}
e^{-i\mathbf{k}\cdot\left(\mathbf{R}_i-\mathbf{R}_l\right)}\cdot\hat{\mathbf{e}} 
\, \hat{\mathbf{e}}
-N^{-1} \hat{\mathbf{e}} \sum_{l} \hat{\mathbf{e}}\cdot \mathcal{H}_{lj}
e^{-i\mathbf{k}\cdot\left(\mathbf{R}_l-\mathbf{R}_j\right)}
\nonumber \\ && 
+ N^{-2}  \hat{\mathbf{e}} \sum_{l,m} 
\hat{\mathbf{e}}\cdot \mathcal{H}_{lm}\cdot\hat{\mathbf{e}}\,
e^{-i\mathbf{k}\cdot\left(\mathbf{R}_l-\mathbf{R}_m\right)}
\, \hat{\mathbf{e}} =
\left[\mathcal{Q} \mathcal{H}(\mathbf{k}=0)\mathcal{Q}\right]_{ij}
+ \left[\delta\mathcal{H}^{\mathcal{Q}1}(\mathbf{k})\right]_{ij}
+ \left[\delta\mathcal{H}^{\mathcal{Q}2}(\mathbf{k})\right]_{ij} + o(k^2),
\end{eqnarray}
where the matrix elements of the terms of the first and second order in $k$, 
$\delta\mathcal{H}^{\mathcal{Q}1}(\mathbf{k})$ and
$\delta\mathcal{H}^{\mathcal{Q}2}(\mathbf{k})$, read
\begin{eqnarray}\label{QHkQ1}
\left[\delta\mathcal{H}^{\mathcal{Q}1}(\mathbf{k})\right]_{ij} 
= i\left(1-\delta_{ij}\right)
\left\{
\frac{\partial^2 V(R_{ij})}{\partial \mathbf{R}_i^2}
\mathbf{k}\cdot\left(\mathbf{R}_i-\mathbf{R}_j\right)
- N^{-1} \sum_{l}
\hat{\mathbf{e}}\cdot\left[\frac{\partial^2 V(R_{il})}{\partial \mathbf{R}_i^2}
\mathbf{k}\cdot\left(\mathbf{R}_i-\mathbf{R}_l\right)
-\frac{\partial^2 V(R_{lj})}{\partial \mathbf{R}_j^2}
\mathbf{k}\cdot\left(\mathbf{R}_j-\mathbf{R}_l\right)\right]\, \hat{\mathbf{e}}\right\},
\nonumber \\
\end{eqnarray}
\begin{eqnarray}\label{QHkQ2}
\left[\delta\mathcal{H}^{\mathcal{Q}2}(\mathbf{k})\right]_{ij} 
&=& \frac{1}{2} \left(1-\delta_{ij}\right)
\frac{\partial^2 V(R_{ij})}{\partial \mathbf{R}_i^2}
\left(\mathbf{k}\cdot\left(\mathbf{R}_i-\mathbf{R}_j\right)\right)^2
- \frac{1}{2}N^{-1} \sum_{l\neq i}\frac{\partial^2 V(R_{il})}{\partial \mathbf{R}_i^2}
\cdot\hat{\mathbf{e}}
\left(\mathbf{k}\cdot\left(\mathbf{R}_i-\mathbf{R}_l\right)\right)^2 
\, \hat{\mathbf{e}}
\nonumber \\ && 
- \frac{1}{2}N^{-1} \hat{\mathbf{e}} \sum_{l\neq j}
\hat{\mathbf{e}}\cdot \frac{\partial^2 V(R_{lj})}{\partial \mathbf{R}_l^2}
\left(\mathbf{k}\cdot\left(\mathbf{R}_l-\mathbf{R}_j\right)\right)^2
+ \frac{1}{2}N^{-2} \hat{\mathbf{e}} \sum_{l\neq m}
\hat{\mathbf{e}}\cdot \frac{\partial^2 V(R_{lm})}{\partial \mathbf{R}_l^2}
\cdot\hat{\mathbf{e}}
\left(\mathbf{k}\cdot\left(\mathbf{R}_l-\mathbf{R}_m\right)\right)^2
\, \hat{\mathbf{e}}.
\nonumber \\
\end{eqnarray}
\end{widetext}

Next, we assume that for small wavevectors $k$ we can treat terms 
$\delta\mathcal{H}^{\mathcal{Q}1}(\mathbf{k})$ and
$\delta\mathcal{H}^{\mathcal{Q}2}(\mathbf{k})$ in the denominator of Eq. (\ref{SEo})
of the main text perturbatively. Due to the symmetry, the term of the 
first order in $k$, $\delta\mathcal{H}^{\mathcal{Q}1}(\mathbf{k})$, 
will contribute in the second order of the perturbation expansion. In contrast,
the term of the second order in $k$, $\delta\mathcal{H}^{\mathcal{Q}2}(\mathbf{k})$, 
will contribute in the first order. Here we will show the contribution of 
$\delta\mathcal{H}^{\mathcal{Q}2}(\mathbf{k})$ term. It reads
\begin{widetext}
\begin{eqnarray}\label{SE2}
&& \delta \boldsymbol{\Sigma}^{\mathcal{Q}2}(\mathbf{k};\omega)
\\ \nonumber &=& 
\!\! - \!\! \sum_{\substack{\text{eigenvec.}\\ p,q}}
\left<1\left|\mathcal{H}(\mathbf{k})\mathcal{Q}\right|\mathcal{E}_p\right>
\left<\mathcal{E}_p
\left|\left(-(\omega+i\epsilon)^2+\lambda_p\right)^{-1}\right|
\mathcal{E}_p\right>
\left<\mathcal{E}_p\left| 
\delta\mathcal{H}^{\mathcal{Q}2}(\mathbf{k})\right|\mathcal{E}_q\right>
\left<\mathcal{E}_q \left|\left(-(\omega+i\epsilon)^2+\lambda_q\right)^{-1}\right|
\mathcal{E}_q\right>
\left<\mathcal{E}_q\left|\mathcal{Q} \mathcal{H}(\mathbf{k})\right|1 \right>.
\end{eqnarray}
\end{widetext}
Counting powers of $k$ in the expression above shows that, at least perturbatively, term 
$\delta\mathcal{H}^{\mathcal{Q}2}(\mathbf{k})$ results in a correction that is
higher order in $k$ than the dominant small wavevector result of approximation 
$\mathcal{H}(\mathbf{k})\approx \mathcal{H}$ in the denominator of Eq. (\ref{SEo}).

\section{Simulation details} 
We obtained zero-temperature glasses by instantaneously quenching supercooled liquids 
of unit number density, $\rho = 1.0$, 
equilibrated through the swap Monte Carlo algorithm \cite{berthier_prx2017}. The 
constituent particles of these liquids have unit mass and diameters $\sigma$ selected 
using distribution $P(\sigma)=\frac{A}{\sigma^3}$, where $\sigma\in [0.73,1.63]$ and 
$A$ is a normalization factor. The cross-diameter $\sigma_{ij}$ is determined 
according to a non-additive mixing rule, 
$\sigma_{ij}=\frac{\sigma_{i}+\sigma_{j}}{2}(1-\epsilon |\sigma_{i}-\sigma_{j}|)$ 
with $\epsilon=0.2$. The interaction between two particles ${i}$ and ${j}$  is 
given by the inverse power law potential,
$ V(r_{ij}) = \left(\sigma_{ij}/r_{ij}\right)^{12} + V_{cut}(r_{ij})$,
when the separation $r_{ij}$ is smaller than the potential cutoff 
$r_{ij}^c=1.25\sigma_{ij}$, and zero otherwise. Here,  
$V_{cut}(r_{ij})=c_{0}+c_{2}\left(r_{ij}/\sigma_{ij}\right)^{2}
+ c_{4}\left(r_{ij}/\sigma_{ij}\right)^{4}$, and  the  coefficients $c_{0}$, 
$c_{2}$ and $c_{4}$ are chosen to guarantee the continuity of $V(r_{ij})$ at $r_{ij}^c$ 
up to the second derivative.

The number of particles $N$ varied between $48000$ and $192000$. The largest systems 
had to be analyzed to determine sound attenuation at the lowest wavevectors reported.

\section{Plane-wave approximation}
We assume that for small wavevectors we can approximate eigenvectors of
the Hessian matrix by plane waves. We note that strictly speaking,
for our amorphous solids the normalization factor is configuration-dependent.
We checked that this dependence is weak and for this reason we use the 
following approximation,
\begin{eqnarray}\label{pwdef}
\mathcal{E}_j^p \approx N^{-1/2} 
\hat{\mathbf{e}}_\mathbf{q} e^{-i\mathbf{q}\cdot\mathbf{R}_j}.
\end{eqnarray}
Approximate plane-wave eigenvectors are labeled by their wavevector $\mathbf{q}$
and their polarization $\hat{\mathbf{e}}_\mathbf{q}$. For each wavevector $\mathbf{q}$
we have one longitudinal and two transverse modes. We assume that the associated
eigenvalues are given by $(v_L q)^2$ and $(v_T q)^2$ for the longitudinal and
transverse modes, respectively. 

Here we will present the derivation of approximate formula for the contribution
to the transverse sound damping coefficient originating from $\Theta$, 
Eq. (\ref{dampingpw1}) of the main text. The contribution originating from 
$\Delta$, Eq. (\ref{dampingpw2}) of
the main text and the approximate expression for the longitudinal sound damping
can be derived in a similar way.

First, we need to calculate
\begin{widetext}
\begin{eqnarray}\label{pw1}
-iN^{-1/2} \sum_j \boldsymbol{\Xi}_{j, \gamma\delta}\hat{e}_\gamma k_\delta\cdot
\boldsymbol{\mathcal{E}}_{j}^p
\approx
-iN^{-1} \sum_j \boldsymbol{\Xi}_{j, \gamma\delta}\hat{e}_\gamma k_\delta\cdot
\hat{\mathbf{e}}_\mathbf{q} e^{-i\mathbf{q}\cdot\mathbf{R}_j} 
= iN^{-1} \sum_j 
\sum_{l\neq j} \frac{\partial^2 V(R_{jl})}{\partial R_{j,\gamma}\partial \mathbf{R}_{j}}
R_{jl,\delta}\hat{e}_\gamma k_\delta\cdot
\hat{\mathbf{e}}_\mathbf{q} e^{-i\mathbf{q}\cdot\mathbf{R}_j}.
\end{eqnarray}
Using the $i\leftrightarrow j$ symmetry we get
\begin{eqnarray}\label{pw2}
&& iN^{-1} \sum_j 
\sum_{l\neq j} \frac{\partial^2 V(R_{jl})}{\partial R_{j,\gamma}\partial \mathbf{R}_{j}}
R_{jl,\delta}\hat{e}_\gamma k_\delta\cdot
\hat{\mathbf{e}}_\mathbf{q} e^{-i\mathbf{q}\cdot\mathbf{R}_j} = 
\frac{i}{2N}\sum_j
\sum_{l\neq j} \frac{\partial^2 V(R_{jl})}{\partial R_{j,\alpha}\partial R_{j,\gamma}}
R_{jl,\delta} \hat{e}_{\mathbf{q}\alpha} \hat{e}_\gamma k_\delta
\left[1-e^{i\mathbf{q}\cdot\left(\mathbf{R}_j-\mathbf{R}_l\right)}\right]
e^{-i\mathbf{q}\cdot\mathbf{R}_j}
\nonumber \\ &=&
\frac{1}{2N}\sum_j
\sum_{l\neq j} \frac{\partial^2 V(R_{jl})}{\partial R_{j,\alpha}\partial R_{j,\gamma}}
R_{jl,\delta}R_{jl,\beta} \hat{e}_{\mathbf{q} \alpha} q_\beta  \hat{e}_\gamma k_\delta
e^{-i\mathbf{q}\cdot\mathbf{R}_j} + o(q) = 
\left(\rho N\right)^{-1} \sum_j 
A^\text{Born}_{j,\alpha\beta\gamma\delta}
\hat{e}_\mathbf{q\alpha} q_\beta \hat{e}_\gamma k_\delta
e^{-i\mathbf{q}\cdot\mathbf{R}_j} + o(q).
\end{eqnarray}
\end{widetext}

Next, we need to take the square of the absolute value of expression (\ref{pw2})
for a given wavevector $\mathbf{q}$ and polarization $\hat{\mathbf{e}}_\mathbf{q}$
and then integrate over spherical shell with frequency 
$q v_L =k v_T$ for longitudinal and
$q v_T =k v_T$ for transverse modes. We shall note that since the spherical shell is 
specified in the frequency space, there will be additional factors, $1/v_L$ 
for longitudinal and $1/v_T$ for transverse modes. Finally, we need to 
multiply the result by $\pi/(2 v_T^2 k^2)$ to get the contribution to the 
transverse sound damping coefficient.

To perform these calculations we assume that wavevector $\mathbf{k}$ is 
parallel to the $y$ axis and the sound polarization $\hat{\mathbf{e}}$ is 
along the $x$ axis. Furthermore, we specify the polarization vector for
the approximate plane-wave eigenvectors as 
$\hat{\mathbf{e}}_\mathbf{q}^\text{L}=\hat{\mathbf{q}}\equiv (\sin\theta\cos\phi, 
\sin\theta\sin\phi,\cos\theta)$ for the longitudinal modes and
$\hat{\mathbf{e}}_\mathbf{q}^\text{T1} = 
(\cos\theta\cos\phi, \cos\theta\sin\phi,-\sin\theta)$ and 
$\hat{\mathbf{e}}_\mathbf{q}^\text{T2} = (-\sin\phi,\cos\phi, 0)$ for the two
transverse modes. 

The contribution of the longitudinal modes reads
\begin{eqnarray}\label{pw3}
&& \frac{\pi}{2 \left(v_T k\right)^2} \frac{v_T^4}{v_L^5} \frac{Vk^6}{(2\pi)^3}
\\ \nonumber \times 
&&
\int d\hat{\mathbf{q}}
\left|\left(\rho N\right)^{-1} \sum_j 
A^\text{Born}_{j,\alpha\beta xy}
\hat{e}^\text{L}_\mathbf{q\alpha} 
\hat{q}_\beta e^{-i\hat{\mathbf{q}}\left(k v_T/v_L\right)
\cdot\mathbf{R}_j}\right|^2.
\end{eqnarray}

Guided by our numerical calculations we assume that the following small wavevector
limit is finite and does not depend on the direction
\begin{eqnarray}\label{pw4}
&& \lim_{\mathbf{q}\to 0} N^{-1}
\sum_j 
A^\text{Born}_{j,\alpha\beta xy}e^{i\mathbf{q}\cdot\mathbf{R}_j}
\sum_l 
A^\text{Born}_{l,\gamma\delta xy}e^{-i\mathbf{q}\cdot\mathbf{R}_l}
\nonumber \\ &\equiv& 
\left<A^\text{Born}_{\alpha\beta xy}A^\text{Born}_{\gamma\delta xy}\right>.
\end{eqnarray}

Expression (\ref{pw3}) becomes
\begin{eqnarray}\label{pw5}
&& \frac{\pi}{2 \left(v_T k\right)^2} \frac{v_T^4}{v_L^5} \frac{Vk^6}{(2\pi)^3}
\frac{1}{\rho^2N}
\left<A^\text{Born}_{\alpha\beta xy}A^\text{Born}_{\gamma\delta xy}\right>
\int d\hat{\mathbf{q}}
\hat{e}^\text{L}_\mathbf{q\alpha} \hat{q}_\beta 
\hat{e}^\text{L}_\mathbf{q\gamma} \hat{q}_\delta
\nonumber \\ &=& 
\frac{\pi}{2 \left(v_T k\right)^2} \frac{v_T^4}{v_L^5} \frac{Vk^6}{(2\pi)^3}
\frac{1}{\rho^2N}
\left<A^\text{Born}_{\alpha\beta xy}A^\text{Born}_{\gamma\delta xy}\right>
\nonumber \\ && \times 
\frac{4\pi}{15}\left(\delta_{\alpha\beta}\delta_{\gamma\delta}
+ \delta_{\alpha\gamma}\delta_{\beta\delta}
+ \delta_{\alpha\delta}\delta_{\beta\gamma}\right)
\nonumber \\ &=& 
\frac{1}{60 \pi} \frac{k^4}{\rho^3} \frac{v_T^2}{v_L^5} 
\left[\left<A^\text{Born}_{\alpha\alpha xy}A^\text{Born}_{\beta\beta xy}\right>
+ \left<A^\text{Born}_{\alpha\beta xy}A^\text{Born}_{\alpha\beta xy}\right>
\right. \nonumber \\ && \left.
+ \left<A^\text{Born}_{\alpha\beta xy}A^\text{Born}_{\beta\alpha xy}\right> \right].
\end{eqnarray}

Assuming again that the small wavevector limit of the correlation
functions of local wave propagation coefficients is finite and does not depend on 
the direction, the contribution of the two transverse modes reads
\begin{eqnarray}\label{pw6}
&& \frac{\pi}{2 \left(v_T k\right)^2} \frac{1}{v_T} \frac{Vk^6}{(2\pi)^3}
\frac{1}{\rho^2N}
\left<A^\text{Born}_{\alpha\beta xy}A^\text{Born}_{\gamma\delta xy}\right>
\nonumber \\ && \times 
\int d\hat{\mathbf{q}} \left(
\hat{e}^\text{T1}_\mathbf{q\alpha} \hat{q}_\beta 
\hat{e}^\text{T1}_\mathbf{q\gamma} \hat{q}_\delta
+ \hat{e}^\text{T2}_\mathbf{q\alpha} \hat{q}_\beta 
\hat{e}^\text{T2}_\mathbf{q\gamma} \hat{q}_\delta \right)
\nonumber \\ &=& 
\frac{\pi}{2 \left(v_T k\right)^2} \frac{1}{v_T} \frac{Vk^6}{(2\pi)^3}
\frac{1}{\rho^2N}
\left<A^\text{Born}_{\alpha\beta xy}A^\text{Born}_{\gamma\delta xy}\right>
\nonumber \\ && \times 
\frac{4\pi}{15}
\left(4\delta_{\alpha\gamma}\delta_{\beta\delta}-
\delta_{\alpha\beta}\delta_{\gamma\delta}
- \delta_{\alpha\delta}\delta_{\beta\gamma}\right)
\nonumber \\ &=& 
\frac{1}{60 \pi} \frac{k^4}{\rho^3}\frac{1}{v_T^3} 
\left[4\left<A^\text{Born}_{\alpha\beta xy}A^\text{Born}_{\alpha\beta xy}\right>
- \left<A^\text{Born}_{\alpha\alpha xy}A^\text{Born}_{\beta\beta xy}\right>
\right. \nonumber \\ && \left.
- \left<A^\text{Born}_{\alpha\beta xy}A^\text{Born}_{\beta\alpha xy}\right> \right]
\end{eqnarray}

Adding expressions (\ref{pw5}) and (\ref{pw6}) we get Eq. (\ref{dampingpw1}) 
of the main text.


\begin{thebibliography}{99}
\bibitem{commentnonz} In $d$ spatial dimensions 
the Hessian matrix has a $d$-degenerate zero eigenvalue 
corresponding to $d$ independent unit vectors; these eigenvalues are 
removed by the orthogonal projection in Eq. (\ref{SEo}).
\end{thebibliography}
\end{document}